\documentclass[conference,a4paper,10pt]{IEEEtran}
\usepackage{CoverPage_IC1004}
\usepackage{graphicx}
\usepackage{pstool}
\usepackage{ctable}
\usepackage{lettrine}
\usepackage{xcomment}
\usepackage{times,amsmath}
\usepackage{amssymb}
\usepackage{amsfonts}
\usepackage [utf8x]{inputenc}
\usepackage[T1]{fontenc}
\usepackage{multirow}
\usepackage{cite}
\usepackage{color}
\usepackage{placeins}
\usepackage{rotating}
\usepackage{trfsigns}
\usepackage{todonotes}

\usepackage{epstopdf}

\usepackage{tikz}
\usetikzlibrary{automata,arrows,shadows,shapes,decorations.pathmorphing,decorations.markings,backgrounds,positioning,fit,shapes.multipart,patterns,mindmap,trees,chains,calc,matrix}
\definecolor{tui-green}{rgb}{0,0.455,0.478}
\definecolor{tui-blue}{rgb}{0,0.2,0.349}
\definecolor{tui-orange}{rgb}{1.0,0.475,0}
\definecolor{tui-lightblue}{rgb}{0.706,0.863,0.863}

\tikzset{
	every node/.style     = {rectangle, draw=tui-blue, rounded corners, text centered, anchor=north, text=tui-blue, text width=4.5cm},
  TextBox/.style        = {minimum height=0.5cm, fill = tui-blue!20},
  TextBoxComment/.style = {minimum height=0.3cm, fill = gray!10, font=\itshape\tiny},
  TextBoxSplit/.style   = {minimum height=1cm,  rectangle split, rectangle split parts=2, rectangle split part fill={tui-blue!20,tui-blue!10}},
	Arrow/.style          = {->, thick}
}

\usepackage[load-configurations=abbreviations,load-configurations=binary]{siunitx}	
\DeclareSIUnit{\mms}{\milli\squaremetre}
\DeclareSIUnit{\inch}{in}
\DeclareSIUnit{\inchs}{in\squared}
\DeclareSIUnit{\mil}{mil}
\DeclareSIUnit{\Msps}{Msps}
\DeclareSIUnit{\Mbps}{Mbps}
\DeclareSIUnit{\LSB}{LSB}
\DeclareSIUnit{\pFS}{\percent FS}
\DeclareSIUnit{\dBc}{\deci\bel c}
\DeclareSIUnit{\dBm}{\deci\bel m}
\DeclareSIUnit{\dBFS}{\deci\bel FS}
\DeclareSIUnit{\dB}{\deci\bel}
\DeclareSIUnit{\dBi}{\deci\bel i}
\DeclareSIUnit{\hex}{0x}
\DeclareSIUnit{\vp}{\volt_{\text{p}}}
\DeclareSIUnit{\vpp}{\volt_{\text{pp}}}
\DeclareSIUnit{\kb}{\kilo\bit}
\DeclareSIUnit{\kB}{\kilo\byte}
\DeclareSIUnit{\MB}{\mega\byte}
\DeclareSIUnit{\GHz}{\giga\hertz}
\DeclareSIUnit{\MHz}{\mega\hertz}

\DeclareSIUnit{\mus}{\micro\second}
\DeclareSIUnit{\ns}{\nano\second}
\DeclareSIUnit{\fs}{\femto\second}

\tdtitle{A Hybrid Channel Model based on WINNER for Vehicle-to-X Application}
\tdauthor{Peter~Große}
\source{Electronic Measurement Research Lab,\\Ilmenau University of Technology,\\ Germany}
\tdnumber{TD(13)07040} %
\location{Ilmenau, Germany} %
\dates{May 28-31, 2013} %
\address{Technische Universität Ilmenau\\FG Elektronische Messtechnik\\PSF 100 565\\D-98684 Ilmenau\\GERMANY} %
\phone{+49 3677 69 1115} %
\fax{+49 3677 69 1113} %
\email{peter.grosse@tu-ilmenau.de} %

\title{A Hybrid Channel Model based on WINNER for Vehicle-to-X Application}%

\author{
\IEEEauthorblockN{Peter Große, Christian Schneider, Gerd Sommerkorn and Reiner Thomä}\\
\IEEEauthorblockA{Electronic Measurement Research Lab\\ Ilmenau University of Technology, Germany\\peter.grosse@tu-ilmenau.de}}

\begin{document}

\makecoverpage
\cleardoublepage
\maketitle

\begin{abstract}
V2V and V2I channel modeling became recently more of interest. To provide realistic radio channels either expensive measurements or complex ray tracing simulations are mostly used. Stochastic channel models are of low complexity but do not offer that deterministic repeatable realism. Based on the WINNER channel model and a simple single path model, a hybrid model has been developed. The concept relies on a layered structure featuring high flexibility and scalability.
\end{abstract}
\vspace{3mm}

\section{Introduction}
\lettrine[lines=2, lhang=0, loversize=0]{R}{ecent} activities within the field of wireless channel measurements and modeling for vehicle-to-vehicle (V2V) or vehicle-to-infrastructure (V2I) scenarios have shown the increased interest for this research area. They are driven by safety and security requirements coming from Intelligent Transportation Systems (ITS) however also increasing demands for high data throughput in vehicle applications. 

Different surveys on V2V propagation channels from \cite{Molisch2009, Boeglen2011, Wang2009} highlight the key challenges to be met for these channels. In \cite{Boeglen2011, Nuckelt2011} various channel models are compared w.r.t. the application within an vehicular ad-hoc networks (VANET) simulator. Whereby some contributions follow the stochastic channel model approach as \cite{Sen2008, Acosta-Marum2007, Renaudin2009} and provide low complexity but not a deterministic repeatable realism. Examples on ray tracing based deterministic modeling can be found in \cite{Reichardt2011,Nuckelt2011, JonathanLedy2012}. Furthermore geometry-based stochastic channel models (GSCM) are introduced for V2V applications in \cite{Karedal2009, Boeglen2011}. Besides other advantages one major key point of these model types is the embedding of basically arbitrary antenna configuration at both sides of the wireless link. A semi-deterministic approach consisting of a ray tracing step followed by the spatial channel model extended (SCME) was proposed in \cite{Boeglen2011, JonathanLedy2012}.

One of the key challenges for V2V and V2I communication is the evaluation of link or system performance under the consideration of different antenna designs at the vehicle itself. Therefore basically only approaches which have the degree of freedom to allow in a flexible manner the embedding of arbitrary antenna pattern are attractive for future research. Such approaches can be found in the group of ray tracing tools or GSCMs. While ray tracing account for high repeatability and high computational complexity, the GSCMs are statistically proven, have low complexity and are well accepted by the research and industry community. 

Currently no attempts are available to extend a cellular system dedicated channel model to V2V applications. But it is necessary to provide a full picture of wireless communications in the context of vehicles. With some limits channel models for cellular applications can be understood or considered for V2I scenarios, e.g. a classical micro cell scenario with base stations (BS) below the rooftop is similar to a V2I scenario, where the infrastructure side of the link is a lamp post or a traffic light. Based on that it would be intuitive to extend or combine a V2I channel to/with a V2V application. 

The paper is organized as follows: In Section \ref{sec:Requirements} the requirements for a V2V and V2I considering channel model are introduced. The concept of the proposed channel model is introduced in Section \ref{sec:Concept} followed by an discussion about the layered model framework in Section \ref{sec:Framework}. Before a short summary is given, the current limitations will be discussed in Section \ref{sec:ProsAndCons}.
\section{Requirements}\label{sec:Requirements}
The combination of a stochastic channel model and simple geometrical aspects leads to more realism in channel modeling \cite{Molisch2009} - the stochastic part provides a \textit{virtual world} consisting of clusters generated based on statistical parameters and the geometry based part supports realistic sequences and transitions.

A proposed concept using that hybrid approach has to be downward compatible to the existing WINNER model \cite{WinnerScm} in sense of functionality. Furthermore, a simple parametrization of complex simulation runs should be realizable hassle-free. 
Due to the focus on V2X applications the model has to support similar channels for related geometrical setups to become more realistic (e.g. two closely separated vehicles communicating to the same base station should \textit{see} similar clusters). Another point changing propagation conditions and/or environments are. Here e.g. simple geometrical mechanisms are required for switching between LoS and nLoS. Since moving vehicles or objects sometimes become strong deterministic scatterers and contribute to the radio channel they have to be taken into account too. 

To support the modeling of scenario transitions and time evolution, as well as time-dependent geometric relations (distances, etc.) of modeled objects, a global time-stamp is needed as an unique reference.

The design also has to consider requirements to support multi-band simulations.

The following enumerations summarize the requirements from a functional and a non-functional point of view. 

\subsection{Functional requirements}
\begin{itemize}
	\item Support for multi-user simulations of V2X scenarios (V2V and typical macro-cell, etc.)
	\item Simulations covering longer periods of time (handle non-stationarity, time evolution)
	\item Hybrid approach: combine a static and dynamic channel part 
	\begin{itemize}
		\item Generation of quasi-static clusters using classic GSCM (reuse of WINNER)
		\item Generation of dynamic clusters by considering moving objects as scatterer
	\end{itemize}
	\item Support for multi-band simulation
\end{itemize}

\subsection{Non-functional requirements}
\begin{itemize}
	\item Scalable concept design allowing more realism than existing solutions
 	\item Modular and extensible software framework
\end{itemize}

\section{Concept}\label{sec:Concept}
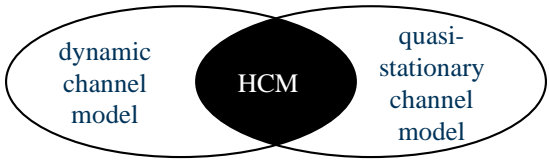
\begin{figure}[!h]
\centering
\tikzset{label/.style={on chain, draw=none, text width=1.9cm, xshift=-1cm}}

\def\firstellipse{(0,-0.65) ellipse (2.3 and 1)}
\def\secondellipse{(2.5,-0.65) ellipse (2.3 and 1)}

\begin{tikzpicture}[start chain=going right]
	\begin{scope}
    	\clip \firstellipse;
    	\fill[fill=black] \secondellipse;
	\end{scope}
    \draw[thick] \firstellipse;
    \draw[thick] \secondellipse;

	\node[label] {dynamic channel model};
	\node[label, text=white] {HCM};
	\node[label] {quasi-stationary channel model};	
\end{tikzpicture}
\caption{Hybrid channel model (HCM)}
\label{fig:conceptHCM}
\end{figure}

To achieve the in the previous section discussed requirements the Hybrid Channel Model (HCM) is proposed. It combines the quasi-stationary channel approach of GSCMs with a geometry based dynamic channel approach. (Figure \ref{fig:conceptHCM}).

The concept proposes a model structure utilizing different aspects of both approaches. The quasi-stationary part considers:
\begin{itemize}
	\item Pairing of stations, channel scenarios
	\item Location-dependent correlated large scale parameters (LSP)
\end{itemize}
The dynamic channel part contributes:
\begin{itemize}
	\item Time-independent geometric context containing environment information
	\item Time-dependent information about all stations (positions, trajectories of moving stations)
\end{itemize}

\begin{figure}[!h]
\centering\input{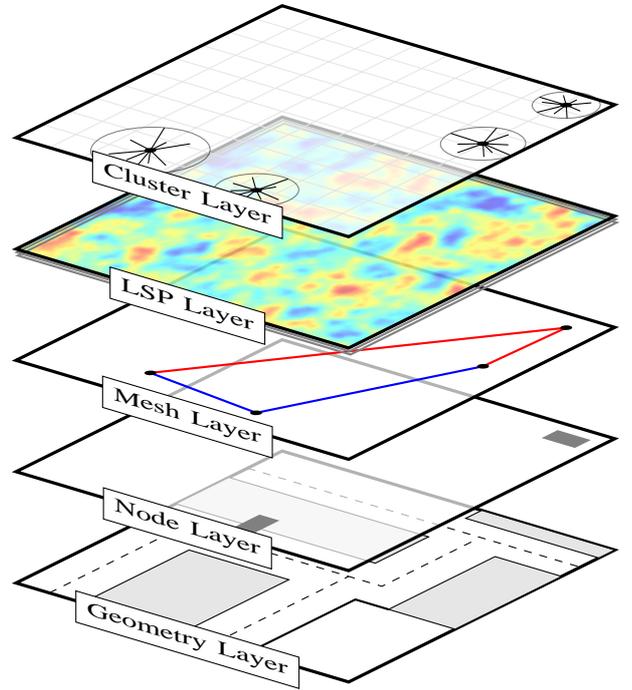}
\caption{Layered model structure of the HCM}
\label{fig:layeredModelStructure}
\end{figure}

Trying to integrate these aspects into existing models would lead to a high-complex monolithic model structure. Therefore the proposed structure is divided into orthogonal layers. The layers are  separated w.r.t. their functionalities and data.
In Figure \ref{fig:layeredModelStructure} the layered model structure of the HCM is shown.

The core element of a channel model is represented by a link between two stations. The description of a link consists of:
\begin{itemize}
	\item Station positions (transmitter and receiver)
	\item Frequency
	\item Channel propagation conditions (channel scenario, LOS/nLOS, fading, shadows, etc.)
\end{itemize}

These link descriptions are modeled as entries in a time-dependent link table as part of the \textit{mesh layer}. (Mesh = set of links between existing stations).

Stations exist and move within a geometric context (or so-called \textit{playground}) modeled as the \textit{geometry layer}, appearing as bottom layer in Figure \ref{fig:layeredModelStructure}. This time-independent layer consists of a set of sub-layer:
\begin{itemize}
	\item Building layer (cuboid object like buildings, used for LOS / nLoS discrimination)
	\item Road layer (surface-bound objects like roads, usable for trajectory generation)
	\item Environment layer (providing classification information e.g. dense urban, to determine channel scenario)
\end{itemize}

On top of the geometry layer the \textit{node layer} contains time-dependent information about all stationary or moving objects used for simulation. In this layer each object is represented by a node, described by:
\begin{itemize}
	\item Node dimension, position or trajectory
	\item Antenna configuration (stations only)
	\item Behavior describing properties (contributes as a scatter, road-bound, etc.)
\end{itemize}
All trajectories are related to a global reference time to support concurrent events.

In geometry-based stochastic channel models large scale parameters are described as stochastic distributions. Using LSP maps location-dependent values of these parameters can be generated independently from the nodes position while maintaining local correlations. These maps are stored within the \textit{LSP layer}, two for each parameter (k-factor, delay spread, angle spread, etc.) due to LOS / nLoS propagation conditions.

As top most layer of the model structure the \textit{cluster layer} contains the location-dependent clusters and propagations properties (delay, path loss, fading, etc.) resulting as output of the hybrid cluster generation.
\section{Framework}\label{sec:Framework}
The proposed model structure is part of a framework describing the structure of a reference implementation of the model. Like the layered model, the framework consists of orthogonal components, separating data and functionality:
\begin{itemize}
	\item Data layers, explained in Section \ref{sec:Concept}
	\item Processing modules, containing the functionality (methods to access/process the data)
	\item Simulation configuration
\end{itemize}

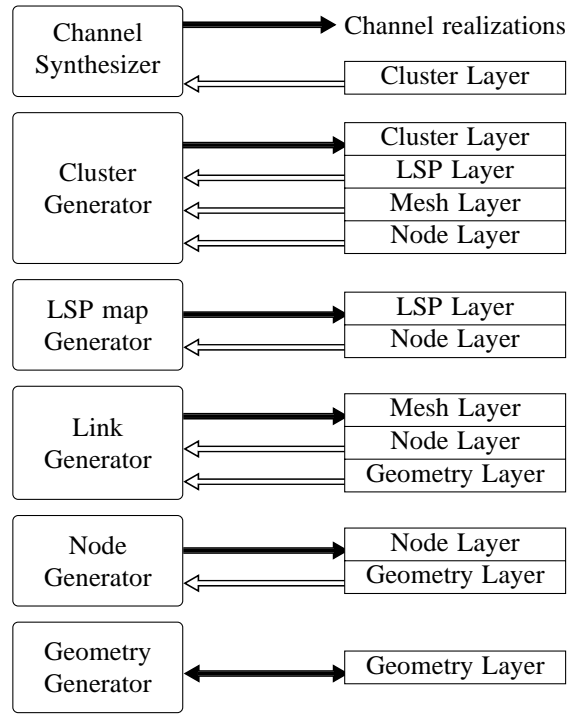
\begin{figure}[!h]
\centering
\begin{tikzpicture}[
	every node/.style={minimum size=1cm},
	node distance=0.2cm,start chain=going below,
	component/.style={rectangle, draw=black, fill=white, align=center, on chain, text width = 2cm, minimum height=1.2cm, rounded corners=2},
	layer/.style={rectangle split, rectangle split parts=1, draw=black, align=center, fill=white, text width = 2.8cm, minimum height=0cm, inner sep=0.15em, xshift=4.5cm},
	vecArrow/.style={semithick,
		decoration={markings,mark=at position 1 with {\arrow[semithick,fill=white]{triangle 60}}},
		double distance=1.3pt, shorten >= 5.5pt, 
		preaction = {decorate},
		postaction = {draw,line width=1.3pt, white,shorten >= 4.5pt}
		},
	vecArrowOutput/.style={semithick,
			decoration={markings,mark=at position 1 with {\arrow[semithick,fill=black]{triangle 60}}}, 
			double distance=1.3pt, shorten >= 5.5pt, 
			preaction = {decorate},
			postaction = {draw,line width=1.3pt, black,shorten >= 4.5pt}
			},
	vecDblArrow/.style={semithick,
		decoration={markings,
			mark=at position 2pt with {\arrowreversed[semithick,fill=black]{triangle 60}},
			mark=at position 1 with {\arrow[semithick,fill=black]{triangle 60}}},
		double distance=1.3pt, shorten >= 5.5pt, shorten <= 5.5pt, 
		preaction = {decorate},
		postaction = {draw,line width=1.3pt, black,shorten >= 4.5pt}
		}
	]
		
   	\node[component] (channel_syn) {Channel \\ Synthesizer};
   	\node[component, minimum height=2cm] (cluster_gen) {Cluster \\ Generator};
   	\node[component] (lspmap_gen)  {LSP map \\ Generator};
   	\node[component, minimum height=1.5cm] (link_gen)    {Link \\ Generator};
   	\node[component] (node_gen)    {Node \\ Generator};
   	\node[component] (geometry_gen){Geometry \\ Generator};
   	
   	\node[layer, right of=channel_syn, yshift=1em, draw=white, text width=3cm](channel){Channel realizations};
   	\node[layer, right of=channel_syn, yshift=-1em] (layer_channel_syn){Cluster Layer};
   	\node[layer, right of=cluster_gen, rectangle split parts=4] (layer_cluster_gen){Cluster Layer	
   		\nodepart{second} LSP Layer
   		\nodepart{third} Mesh Layer
   		\nodepart{fourth} Node Layer};
   	\node[layer, right of=lspmap_gen, rectangle split parts=2] (layer_lspmap_gen){LSP Layer
   		\nodepart{second} Node Layer};
   	\node[layer, right of=link_gen, rectangle split parts=3] (layer_link_gen){Mesh Layer
   		\nodepart{second} Node Layer
   		\nodepart{third} Geometry Layer};
   	\node[layer, right of=node_gen, rectangle split parts=2] (layer_node_gen){Node Layer
   		\nodepart{second} Geometry Layer};
   	\node[layer, right of=geometry_gen] (layer_geometry_gen){Geometry Layer};
   	
	\begin{scope}[on background layer]
		\draw[vecArrowOutput] (channel-|channel_syn) to (channel);
		\draw[vecArrow] (layer_channel_syn.one) to (layer_channel_syn.one-|channel_syn.east);
		
		\draw[vecArrowOutput] (layer_cluster_gen.one-|cluster_gen) to (layer_cluster_gen.one);
		\draw[vecArrow] (layer_cluster_gen.second) to (layer_cluster_gen.second-|cluster_gen.east);
		\draw[vecArrow] (layer_cluster_gen.third) to (layer_cluster_gen.third-|cluster_gen.east);
		\draw[vecArrow] (layer_cluster_gen.fourth) to (layer_cluster_gen.fourth-|cluster_gen.east);
		
		\draw[vecArrowOutput] (lspmap_gen.east|-layer_lspmap_gen.one) to (layer_lspmap_gen.one);
		\draw[vecArrow] (layer_lspmap_gen.second) to (lspmap_gen.east|-layer_lspmap_gen.second);
		
		\draw[vecArrowOutput] (link_gen.east|-layer_link_gen.one) to (layer_link_gen.one);
		\draw[vecArrow] (layer_link_gen.second) to (link_gen.east|-layer_link_gen.second);
		\draw[vecArrow] (layer_link_gen.third) to (link_gen.east|-layer_link_gen.third);
		
		\draw[vecArrowOutput] (node_gen.east|-layer_node_gen.one) to (layer_node_gen.one);
		\draw[vecArrow] (layer_node_gen.second) to (node_gen.east|-layer_node_gen.second);

		\draw[vecDblArrow] (geometry_gen.east|-layer_geometry_gen.one) to (layer_geometry_gen.one);
	\end{scope}
       
\end{tikzpicture}
\caption{Overview of processing modules and the interactions with data layers}
\label{fig:frameworkStructure}
\end{figure}

The concept distinguishes between 6 different types of processing modules (cf. Figure \ref{fig:frameworkStructure}). Each module uses data from different data layers as input and stores it results back into a specific layer.

The \textit{geometry generator} provides the geometric context by filling the mentioned sub-layers. The complexity of this task varies depending on the application scenario from generating only environmental classification data to building detailed virtual environments. Geometries can be randomly generated, user-defined or imported from real-world data (e.g. from open street map).

The \textit{node generator} accomplishes 2 tasks: Node distribution within the geometric context according to the configuration and the generation of trajectories of non-stationary nodes considering the virtual environment (e.g. follow roads).

The \textit{link generator} provides the pairing of the positioned station nodes and assembles the time-dependent link table.

The \textit{LSP map generator} builds maps containing location-dependent values of large scale parameters while considering correlation distances.

The \textit{cluster generator} performs the cluster generation and implements the combination of quasi-stationary and dynamic channel parts. This module also handles time evolution as well as situation transition. The flexible structure allows implementation of further concepts like a cluster memory, enabling the reuse of generated clusters at nearby locations of nodes.

The \textit{channel synthesizer} calculates the actual channel realizations by incorporating the generated clusters and the nodes antenna configuration.

\begin{table}[!h]
\centering
\setlength{\tabcolsep}{0.2em}
\begin{tabular}{p{1.5cm}lcc}
\toprule
Processing \newline Module & Feature & WINNER & V2X \\
\midrule\midrule
Channel		& Antenna embedding 		    & X & X \\
synthesizer	& Sub-path generation	  		& X & X \\
\midrule
Cluster		& Static scatters		& X & X \\
generator	& Dynamic scatters  	&   & X \\
			& Time evolution		& X & X \\
			& Situation transition	&   & X \\
			& Cluster memory	 	&   & X \\
\midrule
LSP map		& LSP map generation	& X & X \\
generator 	& 						&   &   \\
\midrule
Link		& Node pairing (link table assembly)	& X & X \\
generator	& Link state classification 			& X & X \\
			& Detection of reoccurring situations &  & X \\
\midrule
Node		& Node generation (w/ antennas)		& X & X \\
generator	& Node generation (w/o antennas)	&   & X \\
			& Simple movement profiles			& X &  \\
			& Generation of node trajectories	&   & X \\
\midrule
Geometry	& Filling the environ. sub-layer	& X & X \\
generator	& Filling the building sub-layer &   & X \\
			& Filling the road sub-layer 	&   & X \\
			& Random generated environment  & X & X \\
			& User-defined environment      &   & X \\
			& Import of real-world data (geom.) &   & X \\
\bottomrule
\end{tabular}
\caption{Overview of module features and their usage within two example application scenarios}
\label{tab:frameworkModuleFeatures}
\end{table}

Depending on the user-defined simulation configuration single features of the components may be disabled to reduce complexity and limit the computational effort to the required minimum of a task.

The proposed concept can be used in different application scenarios. In Table \ref{tab:frameworkModuleFeatures} the main features of each processing module are outlined and their usage within two example application scenarios are indicated.

Since all parts of the framework interact using generic interfaces, all components are exchangeable and can be replaced by implementations with more advanced features.

%
%
%

\section{Current Limitations}\label{sec:ProsAndCons}

Besides the novel proposal for a scalable, V2V and V2I combining channel model several limitations and research works are identifiable. Some of them will be step-wise solved by the continuous work on the model itself, others need deeper research. Within the contribution of \cite{Karedal2009} dense multi-path contributions are described. This propagation phenomenon is currently not considered within HCM. Furthermore it is not taken into account that several links could be influenced by the same clusters, which in fact will influence the inter-link correlation. Another important issue is a wider discussion on multi-band channel modeling, where correlated clusters may appear in different frequency bands or at least their large scale parameters show a similar characteristic. An open question is, if it would be necessary to model the size of a cluster and if it would be how to model it. This could be interesting in particular for V2V channels, where different car sizes could cause larger or smaller clusters. 
	
	
	
\section{Summary}\label{sec:Summary}
In this paper a new hybrid channel model evolving the WINNER model to be applicable for V2V and V2I system evaluations is proposed. The hybrid characteristic is understood by separating the modeling task into a quasi-static and dynamic part, where the latter considers necessary extensions e.g. for moving clusters (vehicles). The novel layered concept ensures a scalable and flexible GSCM approach. This includes an user-defined level of randomness (at each layer) including user movement modeling, environment descriptions and more. However the concept is down scalable to the WINNER approach due to maintaining its functional parity. By introducing a modular software structure (orthogonal design of layers) the approach is extensible and easy to use.



\bibliographystyle{IEEEtran}
\bibliography{./References_v2x,./References}

\end{document}